\documentclass[graybox]{svmult}

\usepackage{type1cm}        
\usepackage{makeidx}         
\usepackage{graphicx}        
\usepackage{multicol}        
\usepackage[bottom]{footmisc}

\usepackage{newtxtext}       %
\usepackage{newtxmath}       

\usepackage{subcaption}
\usepackage{amsmath}
\usepackage[inline]{enumitem}

\makeindex             

\usepackage{arxiv}

\usepackage[utf8]{inputenc} 
\usepackage[T1]{fontenc}    
\usepackage{hyperref}       
\usepackage{url}            
\usepackage{booktabs}       
\usepackage{amsfonts}       
\usepackage{nicefrac}       
\usepackage{microtype}      
\usepackage{lipsum}
\usepackage{graphicx}
\graphicspath{ {./images/} }

\title{Graph and Network Theory for the analysis of Criminal Networks}
\titlerunning{Graph Theory for the analysis of Criminal Networks}

\author{
 Lucia Cavallaro \\
  University of Derby\\
  Derby, UK \\
  \texttt{l.cavallaro@derby.ac.uk} \\
   \And
 Ovidiu Bagdasar \\
  University of Derby\\
  Derby, UK \\
  \texttt{O.Bagdasar@derby.ac.uk} \\
  \And
 Pasquale De Meo\\
  University of Messina\\ Messina, Italy\\
  \texttt{pdemeo@unime.it} \\
  \And
 Giacomo Fiumara \\
  University of Messina\\ Messina, Italy\\
  \texttt{gfiumara@unime.it} \\
  \And
  Antonio Liotta \\
  Free University of Bozen-Bolzano\\ 
  Bolzano, Italy \\
  \texttt{Antonio.Liotta@unibz.it} \\
}
\authorrunning{L. Cavallaro, O. Bagdasar, P. De Meo, G. Fiumara, and A. Liotta} 

\begin{document}
\maketitle
\begin{abstract}
Social Network Analysis is the use of Network and Graph Theory to study social phenomena, which was found to be highly relevant in areas like Criminology. This chapter provides an overview of key methods and tools that may be used for the analysis of criminal networks, which are presented in a real-world case study. Starting from available juridical acts, we have extracted data on the interactions among suspects within two Sicilian Mafia clans, obtaining two weighted undirected graphs. Then, we have investigated the roles of these weights on the criminal network's properties, focusing on two key features: weight distribution and shortest path length. We also present an experiment that aims to construct an artificial network that mirrors criminal behaviours. To this end, we have conducted a comparative degree distribution analysis between the real criminal networks, using some of the most popular artificial network models: Watts-Strogatz, Erd\H{o}s-R\'{e}nyi, and Barab\'{a}si-Albert, with some topology variations. This chapter will be a valuable tool for researchers who wish to employ social network analysis within their own area of interest.
\end{abstract}

\section*{Introduction}
\label{sec:intro}
Graph Theory is a well established field in mathematics. However, only recently many of its theoretical results started to be used within Social Network Analysis (SNA), an area with significant implications for real world scenarios. For example, one can simulate the behaviour of social networks using strategies like link predictions~\cite{LU20111150, Hasan2011}, temporal networks, or spreading of influences~\cite{Tassiulas2013,Cavallaro2020ArXiv}. Other practical applications include to deal with large Artificial Neural Networks~\cite{Mocanu2018network,Cavallaro2019network,Cavallaro2020network} or targeted advertisements to people based on their friends' interests~\cite{Bellur2007} or, on the other side, containing the spread of fake news~\cite{zhou2018fake}. 

Network Science tools may also be used in the investigation of criminal networks. Sometimes the complex social interactions within a clan-based society may help the feature selection process for building machine learning models~\cite{Oluwabunmi2019IEEE}. Other times, it is Network Science itself that helps conducting better performing investigation from law enforcement agencies. To this end, criminal networks can be encoded as graphs, and various types of analysis and simulations can be carried out for modelling criminal behaviours.

This chapter is to intended as a short tutorial on how Network Science strategies may be used to conduct an in-depth analysis on real criminal networks. Here, the Sicilian Mafia scenario has been considered. Sect.~\ref{sec:mafia} relates to our previous analysis on this topic. In particular, Sect.~\ref{subsec:degdistr} includes a comparative Degree Distribution analysis between real criminal networks and artificial ones. Indeed, when it is possible to find out a synthetic network reflecting the behaviour of real-world criminal networks, law enforcement agencies (LEAs) and network scientists can recreate those networks and simulate how interconnections among criminal will evolve.

This chapter is structured as follows. Sect.~\ref{sec:background} presents the key theoretical tools (required for understanding the experiments conducted in Sect.~\ref{sec:mafia}), and it is divided in two parts:
\begin{enumerate*}[label=(\roman*)] 
\item tools, where the basic definitions on network science are provided; and
\item popular artificial networks description (as the topologies used in Sect.~\ref{subsec:degdistr}).
\end{enumerate*}
Next, in Sect.~\ref{sec:sna} a brief review on the use of \begin{enumerate*}[label=(\roman*)] \item Social Network Analysis, and its implication in \item Criminal Networks is defined. \end{enumerate*}
Sect.~\ref{sec:mafia} is a case study summarizing our work on two real criminal networks related to Sicilian Mafia~\cite{Ficara2020, cavallaro2020disrupting}. This section includes four parts:
\begin{enumerate*}[label=(\roman*)] 
\item datasets description, based on the data extracted from juridical acts;
\item weights distribution analysis, which represents an important preliminary study to understand how the interactions among suspected are structured in terms of interaction frequency; 
\item shortest path analysis, which allows to identify trusted affiliates inside the clan who can spread confidential and illegal messages;
\item comparative degree distribution analysis between real and synthetic networks, to artificially recreate the criminal networks used here, with the purpose of conducting further investigations through them.
\end{enumerate*}
Finally, the conclusions follow in Sect.~\ref{sec:conclusions}.

\section*{Complex Networks}
\label{sec:background}

In this section we introduce the main Network Science concepts underpinning Social Network Analysis, which are later exemplified in the criminal network case study (Sect.~\ref{sec:mafia}). In particular, in Sect.~\ref{subsec:tools} the main definitions required for understanding the mechanics of SNA are provided. All theoretical concepts are derived from~\cite{barabasi2016network}, which we refer to the reader for further technical details. 

\subsection{Tools}
\label{subsec:tools}
In this section, we start with some basic definitions of Graph Theory.
\begin{trailer}{Graph}
\begin{definition}
A \textit{graph} denoted by $G = (N, E)$, consists of a set of nodes $N$ and a set of edges $E \subseteq N \times N$ (also called links $L$). It is a convenient way of representing relationships between pairs of objects.
\end{definition}

As an example, \textit{Facebook\textregistered} may be viewed as a graph, where the nodes represent users and edges represent the friendship relationship among them. Is it also possible to define a \textit{subgraph} as follows:

\begin{definition}
A \textit{subgraph} $H$ of the graph $G$ is a graph whose nodes and edges are subsets of the nodes and edges of $G$.
\end{definition}

Furthermore, graphs may be either \textit{weighted}, or \textit{unweighted}:

\begin{definition}
A {\em weighted graph} $G = (N, E, W)$ is a triplet consisting of a finite set of nodes $N$, a set of edges $E$, and a set of weights $W: E \rightarrow \mathbb{R}$ defined on each edge. If all edges weights are equal to one, then the graph is called {\em unweighted}.
\end{definition}
\end{trailer}

\begin{trailer}{Degree} 
\begin{definition}
The \textit{degree} of a node $n_i$, denoted $deg(i)$ or $k_i$, is the number of incident edges to $n_i$.
The sum of the degrees of all nodes is equal to the double of the number of edges $E$:
\begin{equation}
    \sum_{n \: \in \: N}k_n = 2E.
\end{equation}
\end{definition}

\begin{definition}
In weighted networks, the {\em weighted degree} (also known as {\it strength} \cite{antoniou2008statistical,barrat2005}) is the sum of the edges weights $w$ incident on $n_i$:
\begin{equation}
    k_i = \sum_{(i,j) \in E} w_{ij},
\end{equation}
where the summation spans over all edges $(i,j)$ in the network, linked to node $n_i$.
 \end{definition}

\begin{definition}
For undirected networks, the \textit{average degree} is defined as

\begin{equation}
    \label{eq:ave-deg}
    \langle k \rangle = \frac{1}{N}\sum_{i=1}^N k_i = \frac{2L}{N},
\end{equation}
where $N$ is the total number of nodes, $k_i$ is the degree of a generic node $i$, and $L$ represents the total number of links, or edges $E$, within the network.

\begin{svgraybox}
\paragraph{Small-World}
The Small World phenomenon~\cite{smallworld1967,smallworld1969} is based on the concept of the six degrees of separations, according to which two random people in the world may be connected each other via a few acquaintances (i.e., it is estimated that there are six people in the middle between the source and the destination). 
In Network Science, it translates into a ``short'' distance between two randomly chosen nodes within a network, that is
\begin{equation}
    \label{eq:smallworld}
    \langle d \rangle \approx \frac{\ln N}{\ln \, \langle k \rangle},
\end{equation}
where $N$ is the total number of nodes in the graph, $\langle k \rangle$ is the network average degree, and $\langle d \rangle$ the average distance within the network. The denominator implies that the denser the network, the smaller the distance between the nodes is.
In conclusion, the average path length or the diameter depends logarithmically on the system size.
\end{svgraybox}
\end{definition}

\subparagraph{Degree Distribution}
The {\it degree distribution} $p_k$ provides the probability that a randomly selected node in the network has degree $k$. Since $p_k$ is a probability, it must be normalized; i.e.,
\begin{equation}
    \sum_{i=1}^{\infty}p_k = 1.
\end{equation}

For a network made of $N$ nodes, the degree distribution is the normalized histogram given by:
\begin{equation}
p_k = \frac{N_k}{N},
\end{equation}
where $N_k$ is the number of nodes having degree $k$.

The degree distribution has assumed a central role in network theory following the discovery of scale-free networks (See ``Scale-Free Property'' paragraph); moreover, $p_k$ determines many network phenomena, from network robustness to the spread of viruses.

\subparagraph{Weight Distribution} 

The degree distribution can be extended to weighted networks considering the weighted degree (strength) distribution $P(s)$, defined as the probability that a node may have weighted degree (strength) equal to $s$. 
Based on \cite{barrat2005}, this is 
\begin{equation}
P(s) \sim s^{-\gamma},
\end{equation}
where $\gamma$ is a constant typical of the network.

\end{trailer}

\begin{trailer}{Clustering Coefficient}
Clustering is used to quantify the relationship among nodes' neighbours. Indeed, the degree only considers the number of direct links between nodes. The \textit{clustering coefficient} $C_i$ measures the edge density in the immediate neighbourhood of a node. $C_i\in[0,1]$ represents the clustering coefficient of a generic node $n_i$:
\[
\begin{cases} $if $ C_i = 0 , & $there are no edges among the node's neighbours$ \\
$if $ C_i = 1 , & $ each node's neighbour is connected with the others$
\end{cases}
\]
The local clustering coefficient is computed as follow:
\begin{equation}
    \label{eq:cluster}
    C_i = \frac{2L_i}{k_i(k_i-1)},
\end{equation}
where $k_i$ is the degree of the generic node $n_i$, and $L_i$ represents the number of links (i.e., edges) between the $k_i$ neighbours of $n_i$. 

\subparagraph{Average Clustering Coefficient}
The average $\langle C \rangle$ of $C_i\in i=1,\dots,N$ in the whole network is given by
\begin{equation}
\label{eq:ave-cluster}
\langle C \rangle = \frac{1}{N}\sum_{i=1}^N C_i.
\end{equation}
\end{trailer}

\begin{trailer}{Adjacency Matrix} 
A common way to represent relationships among nodes is the \textit{adjacency matrix} $A$.
\begin{definition}
The \textit{Adjacency Matrix} $A[i,j]$ holds node degree (weighted or unweighted) to the edge $(n_i,n_j)$ if it exists, where $n_i$ is the node with index $i$ and $n_j$ is the node with index $j$. If there is no such edge, then $A[i,j]=None$. 
\end{definition}
For undirected graphs $A$ is symmetric (i.e., $A[i,j]=A[j,i]$ $ \forall n_i,n_j \in N$).
\end{trailer}

\begin{trailer}{Path}
 \begin{definition}
 A \textit{path} is a sequence of alternating nodes and edges that flow from a starting node to an ending one such that each edge is incident to its predecessor and successor node.
 A path is called \textit{simple} if each node in the path is distinct.
 \end{definition}
More formally, a path can be defined as a sequence of nodes
$$
P = (n_1, n_2, \dots, n_m) \in N \times N \times \cdots \times N,
$$
such that $n_i$ is adjacent to $n_{i+1}$ for $1 \le i \le m-1$.
Such a path $P$ is called a path of length $m-1$ from $n_1$ to $n_m$.
 
Measures based on paths strategies are the \textbf{shortest path length analysis}.

\end{trailer}

\begin{trailer}{Distance}
\begin{definition}
The \textit{distance} from a node $n_i$ to a node $n_j$ in $G$, denoted $d(n_i,n_j)$ is the length of a \textbf{shortest path} from $n_i$ to $n_j$ (if such a path exists).
$$
d_{ij} = min \left( \Gamma(i,j) \right),
$$
where $\Gamma(i,j)$ is the set of paths connecting $i$ and $j$.

\end{definition}
\end{trailer}

\begin{trailer}{Connectedness}
\begin{definition}
A graph $G$ is \textit{connected} if, for any two nodes, there is a path between them. \end{definition}
\begin{definition}
If $G$ is not connected, its maximal connected subgraphs are called the \textit{connected components} of $G$. 
\end{definition} 
\begin{definition}
If a network consists of two components, a properly placed link can connect them, making the network connected. Such a link is called \textit{bridge}.
\end{definition}
\end{trailer}

\begin{trailer}{Scale-Free Property}
 The majority of real networks, such as the World Wide Web, are called {\it scale-free networks} and follow the definition: \begin{definition}
A scale-free network is a network whose degree distribution follows a power law.
\end{definition} 
The \textit{power-law distribution} has the following form
 \begin{equation}
     p_k \sim k^{-\gamma}, 
 \end{equation}
where the exponent $\gamma$ is its {\it degree exponent}.

Some artificial network models such as the \textbf{Barab\'{a}si-Albert (BA) Model} successfully exhibit this feature.
\end{trailer}

\subsection{Artificial Networks}
\label{subsec:artificial}
The need for scientists to create Artificial, or Synthetic Networks has been born from the aim to reproduce real network properties in a controlled environment. For this reason, several typologies of Artificial Networks have been formulated. 

Three models in particular have found special popularity within the scientific community: The Erd\H{o}s-R\'{e}nyi (ER, also known as Random Network) Model, the Watts-Strogats Model (WS; i.e., a Random Network variation), and the Barab\'{a}si-Albert (BA) Model. 
This last one tries to capture two important properties of real network: the growth and the preferential attachment. Further details on those models are provided in the following paragraphs.
\begin{trailer}{Random Network Model}
A random network consists of $N$ nodes where each node pair $(n_i,n_j), \forall i,j \in N$ is connected with probability $p$. To construct a random network one needs to
\begin{enumerate}
\item Start with $N$ isolated nodes,
\item Select a node pair $(n_i,n_j)$ and generate a random number $rand\in[0,1]$: 
\[
\begin{cases} $if $rand>p , & $connect the selected node pair with a link$ \\
$otherwise$, & $leave them disconnected$
\end{cases}
\]
\item Repeat the previous step for all pairs of distinct nodes $(n_i,n_j) \in N\times N$.
\end{enumerate}

The network obtained after this procedure is called a {\it random graph} or a {\it random network}. 
There are two definitions of a random network: the definition provided in the Erd\H{o}s-R\'{e}nyi Model, and the one of the Gilbert Model.

\begin{svgraybox}
\paragraph{\textit{Erd\H{o}s-R\'{e}nyi} Model} 
Random networks are also called {\it Erd\H{o}s-R\'{e}nyi Networks} from the names of the mathematicians Paul Erd\H{o}s (1913-1996) and Alfr\'{e}d R\'{e}nyi (1921-1970), who studied the properties of these networks. Their model follow the structure
\begin{equation}
\label{eq:er}
G(N, L),
\end{equation}
where $N$ labeled nodes are connected with $L$ randomly placed links (i.e., edges). Paul Erd\H{o}s and Alfr\'{e}d R\'{e}nyi used this definition in their paper~\cite{random1959}.
\end{svgraybox}

\begin{svgraybox}
\paragraph{\textit{Gilbert} Model}
It is a variation of the Erd\H{o}s-R\'{e}nyi Model. It has been defined by Edgar Nelson Gilbert (1923-2013) and follows the structure
\begin{equation}
\label{eq:gr}
G(N, p), 
\end{equation}
where each pair of $N$ labeled nodes is connected with probability $p$.
\end{svgraybox}
\end{trailer}

There are two main limits in Random Network Model that had to be overcome over the years by the academic community: 
\begin{enumerate}
\item The local clustering coefficient in ER model is given by \cite{barabasi2016network}
$$
C_i = \frac{\langle k \rangle}{N}.
$$
This behaviour of $C_i$ is contradicted by the local clustering coefficient of real networks.
\item The Poisson distribution that describes the degree distribution of ER networks does not allow large differences between the worst- and best-connected nodes in the network. 
This implies that hubs, frequently observed in real networks, cannot be found in ER networks. 
BA model, relying on preferential attachment and growth, successfully reproduces this fundamental feature.
\end{enumerate}
In the following paragraphs those models are described.

\begin{trailer}{Watts-Strogats Model}
Two main considerations motivated Duncan J. Watts (1971) and Steven Strogatz (1959) to propose this model:
\begin{enumerate*}[label=(\roman*)] 
\item in real networks the average distance between two nodes depends logarithmically on $N$ (See ``Small-World'' average distance $\langle d \rangle$); 
\item the average clustering coefficient $\langle C \rangle$ of real networks is much higher than expected for a random network of similar $N$ and $L$ (i.e., $E$).
\end{enumerate*}

To construct a random network according to \textit{Watts-Strogats} Model~\cite{watts1998}:
\begin{enumerate}
\item Start from a ring of $N$ nodes, whereas each node is connected to its immediate previous and next neighbours; hence, each node has $\langle C \rangle = 3/4$, initially.
\item With probability $p\in [0,1]$ each link is rewired to a randomly chosen node
\[
\begin{cases} $if $p\simeq 0 , & $regular lattice$ \\
$if $0<p<1 , & $Small-World property$\\
$if $p=1 , & $Random Network Model (all links rewired).$
\end{cases}
\]
\end{enumerate}
 The Watts-Strogatz model interpolates between a {\it regular lattice}, which has high clustering (but lacks the Small-World phenomenon), and a {\it random network}, which has low clustering (but displays the Small-World property). 
 Moreover, high nodes degrees are absent from Watts-Strogatz model.
\end{trailer}

\begin{trailer}{Barab\'{a}si-Albert Model}
This model was theorized by Albert-L\'{a}szl\'{o} Barab\'{a}si (1967) and R\'{e}ka Albert (1972)~\cite{barabasi1999}. 
It simulates a Scale-Free Network rather than a Random one, by introducing two new concepts to the model: network growth and the preferential attachment. 
The first concept assumes that real networks continuously increase over time, so new nodes must be considered. 
The second point argues that random connections defined by a fixed probability do not reflect the behaviour of real networks; in fact, in real scenarios, nodes tend to link to the more connected nodes.
\begin{definition}
The \textit{Preferential Attachment} is the probability $\Pi(k)$ that a link of the new node $n_j$ connects to node $n_i$ depends on the degree $k_i$ through the formula
        \begin{equation}
            \label{eq:pref-att}
            \Pi(k_i) = \frac{k_i}{\sum_j k_j}.
        \end{equation}
\end{definition}

To construct an artificial network with the Barab\'{a}si-Albert model, the steps are:
\begin{enumerate}
    \item Start with a set of $N_0$ nodes, the links between which are chosen arbitrarily, as long as each node has at least one link. 
    \item {\bf Growth} -- At each timestep a new node $n_j$ with $l$ links (with $l\le l_0$) that connects the new node to nodes already in the network is added.
    \item The connections between the new node with the older nodes are defined by the \textbf{Preferential Attachment} probability.
\end{enumerate}{}
\end{trailer}

\section*{Social Network Analysis in Criminal Networks}
\label{sec:sna}
In this section we provide an overview of state-of-the-art of Social Network Analysis applied to Criminal Networks. We also consider the most relevant studies concerning specifically the Sicilian Mafia criminal topologies.

\subsection{Criminal Networks Analysis}
\label{subsec:crime}
Through SNA, LEAs are able to analyze criminal networks and investigate the relations among criminals. For this reason, nowadays there is a growing interest in the application of Graph and Network Science onto criminal networks. For instance,  SNA has been used in~\cite{Chen2004} to build crime prevention systems. However, due to the lack of data availability on those kind of networks, there are difficulties in finding relevant quantitative studies. Such examples are those conducted by Szymanski~\cite{Szymanski2018} and Berlusconi~\cite{BerlusconiPLOS2016}, on the problem of community detection and link prediction.

\subsection{Sicilian Mafia networks}
\label{subsec:literature-mafia}
Sicilian Mafia has a particular structure that differs from common criminal networks (such as the terrorist nets), whereby it is a common practice for criminals to come together to achieve a common goal and then fall apart. By contrast, in Sicilian Mafia this behaviour does not occur. Indeed, the affiliates are bound by blind loyalty and they still pursue further goals even after achieving a previous one. Moreover, Families last for several generations. They also tend to diversify their objectives: from controlling entire economic sectors (e.g., by giving ``protection'' to small traders and taking control of larger factories), to influencing countries political life (e.g., by interfering in the results of electoral competitions). Sicilian Mafia originated in Sicily, and has now spread worldwide~\cite{franchetti1925sicilia, McGloin2005, Mastrobuoni2012}.
The blind loyalty of affiliates makes it even more difficult to obtain reliable information about those criminal networks topologies: important information about such criminal network is likely to be missing or hidden, due to the covert and stealthy nature of criminal actions~\cite{krebs2002,Xu2005,Calderoni2010,Campana2012}.

\section*{Case study: The Sicilian Mafia}
\label{sec:mafia}
This section describes firstly the real criminal datasets we used for our tests, followed by a brief summary on the strategies conducted jointly with the results obtained so far as an example on Network Science strength, and how it can be used to significantly help LEAs. In particular, the experiments relate to: \begin{enumerate*}[label=(\roman*)] 
    \item weight distribution,
    \item shortest path length, and
    \item degree distribution.
\end{enumerate*}{}

\subsection{Dataset Description}
\label{subsec:datset}
The case study example relates to two real-world datasets we built from juridical acts\footnote{Datsets and source code are available at \url{https://github.com/lcucav/criminal-nets}}~\cite{cavallaro2020disrupting}:
\begin{enumerate*}[label=(\roman*)] 
\item the {\em Meetings} dataset represents the physical meetings among criminals obtained through LEA evensdropping;
\item the {\em Phone Calls} dataset refers to phone calls between individuals obtained through LEA interceptions.
\end{enumerate*}

This particular investigation was a prominent operation conducted during the first decade of the 2000s and focused on two Mafia clans known as the ``Mistretta'' family, and the ``Batanesi''clan~\cite{online-journ2019, Ficara2020, cavallaro2020disrupting}.

\begin{figure*}[htb]
  \centering  
  \begin{subfigure}{.49\textwidth}
    \includegraphics[width=\textwidth]{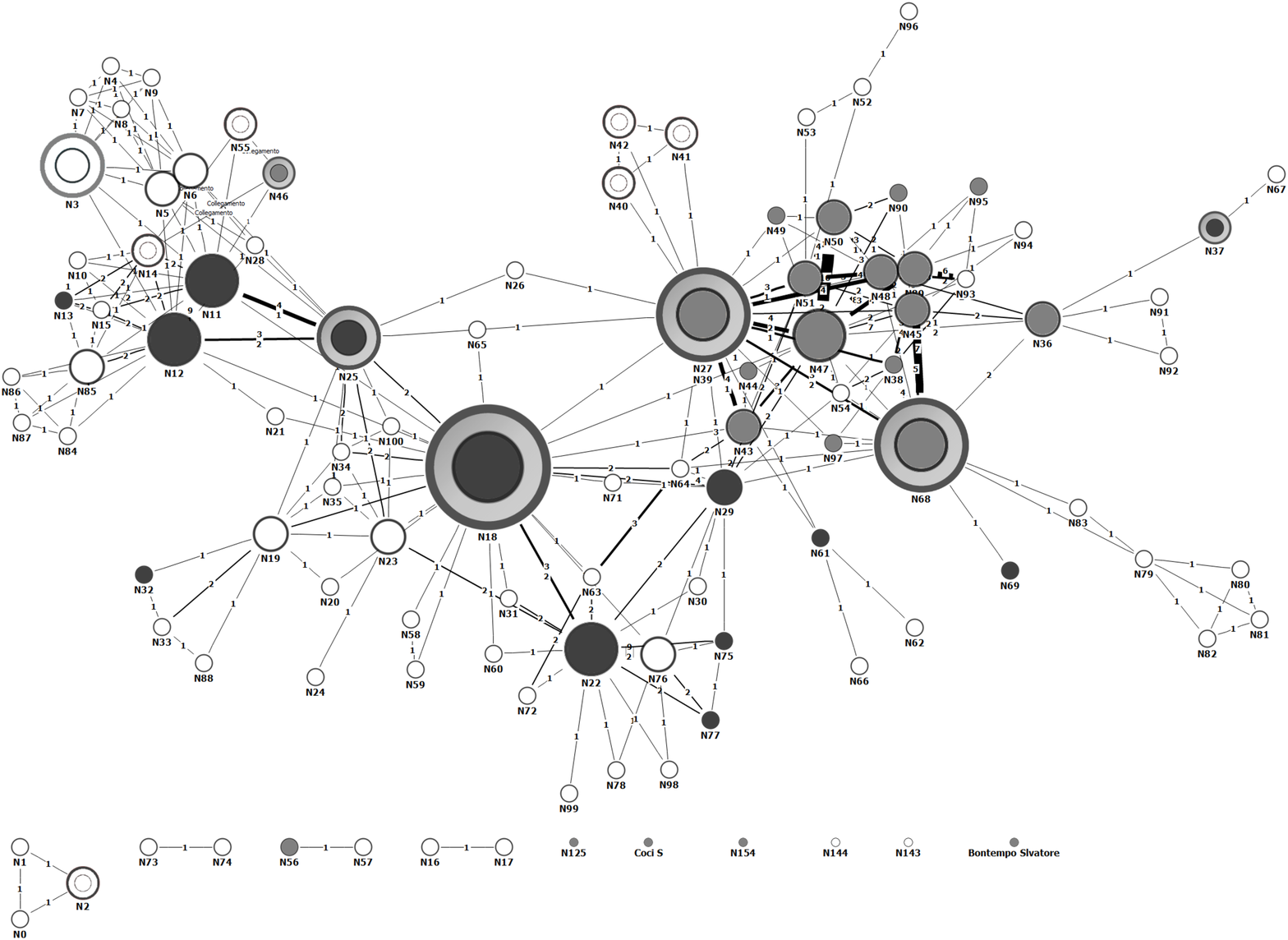}
    \vspace{-2em}
    \caption{\textsc{\textsc{Meetings}}}
  \end{subfigure}
  \begin{subfigure}{.49\textwidth}
    \includegraphics[width=\textwidth]{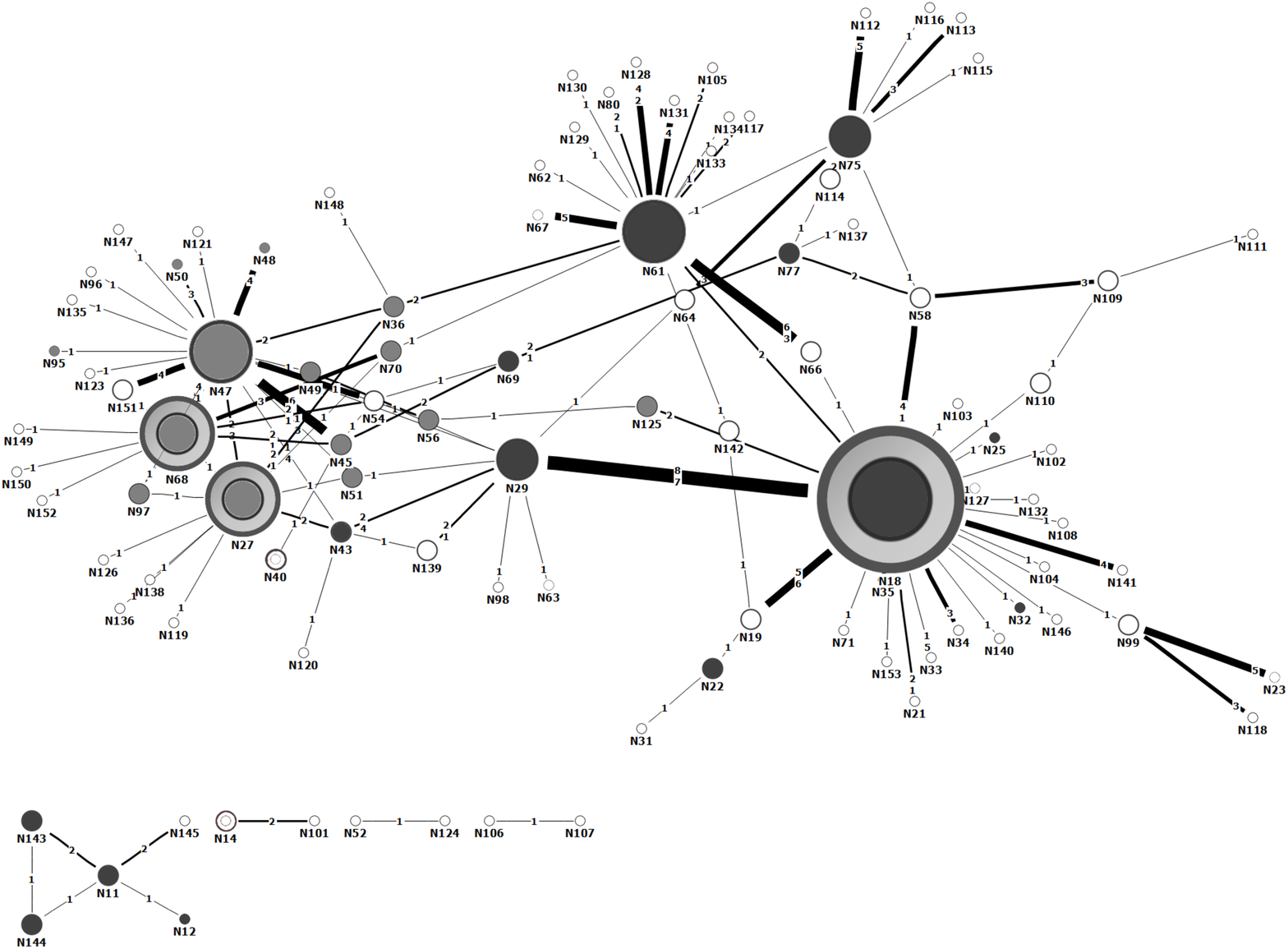}
    \vspace{-2em}
    \caption{\textsc{Phone Calls}}
  \end{subfigure}
    \vspace{-0.8em}
     \caption{Dataset Description. The colours represent different clans: darker nodes are the ``Mistretta'' family; in grey the ``Batanesi'' clan is drawn; white and light gray circled nodes and for two others Mafia families not directly involved in the current investigation. All {\em circled} nodes represent the {\em bosses}. Lastly, white nodes represent other subjects not classifiable in any of the previous categories.
     Edges' width depends on the number of meetings or phone calls, while the nodes size relates with their degree. (Reproduced from Ficara 
     \emph{et al.} 2020)}
\label{fig:graphs}
\end{figure*}

Both datasets led to undirected and weighted graphs; thus, edge weights $w$ are available and represent the number of times any given pair had a meeting in the {\em Meetings} dataset, and the number of times two individuals called each other in {\em Phone Calls} dataset. In SNA, those coefficients are also known as the strength of the tie binding two individuals~\cite{Ficara2020}. In Fig.~\ref{fig:graphs}, the graphs obtained as well as the description of what each element represents (i.e., nodes, colours, edges weight, nodes and edges size, etc.) is shown. The main characteristics of the datasets are summarized in Table~\ref{table:networks}.

\begin{table}[!t]
\caption{Characteristics of {\em Meetings} and {\em Phone Calls} networks. (Reproduced from Ficara \emph{et al.} 2020)}
\label{table:networks}
\begin{tabular}{p{3cm}p{2cm}p{2cm}}
\hline\noalign{\smallskip}
\textbf{Parameter} & \textit{Meetings} & \textit{Phone Calls} \\
\noalign{\smallskip}\svhline\noalign{\smallskip}
No. Nodes & 101 &  100\\ 
No. Edges & 256  & 124\\ 
Max. Weight & 10 & 8\\ 
Max. Frequency & 200 & 100\\ 
Avg. Degree & 5.07 & 2.48\\ 
Max. Shortest Path & 7 & 14\\  \hline
Common nodes & \multicolumn{2}{c}{47} \\ 
\noalign{\smallskip}\hline\noalign{\smallskip}
\end{tabular}
\end{table}

\subsection{Weight Distribution Analysis}
\label{subsec:wdistr}

\begin{figure*}[htb]
  \centering  
  \begin{subfigure}{.4\textwidth}
    \includegraphics[width=\textwidth]{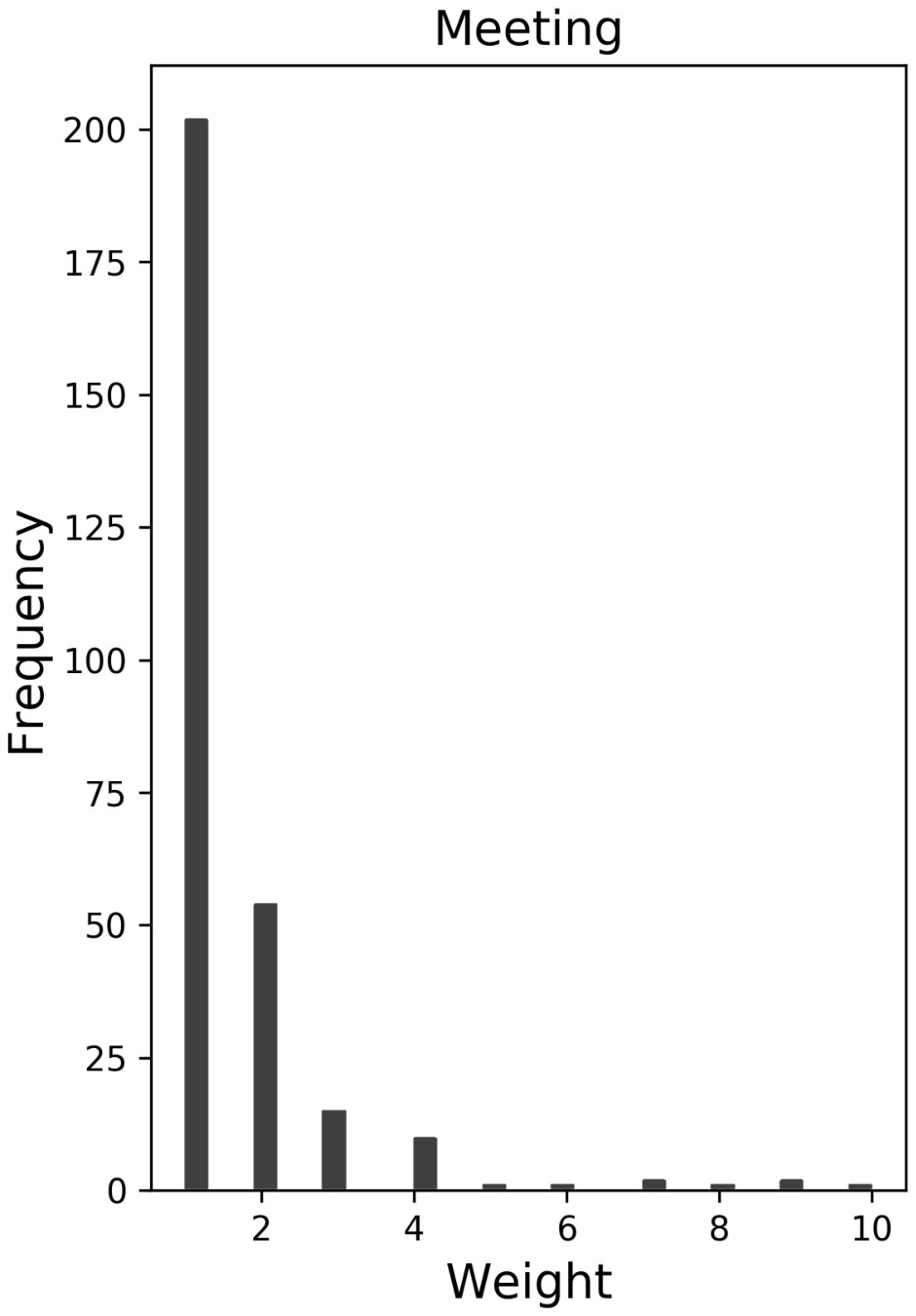}
    \vspace{-2em}
    \caption{\textsc{\textsc{Meetings}}}
  \end{subfigure}
  \begin{subfigure}{.4\textwidth}
    \includegraphics[width=\textwidth]{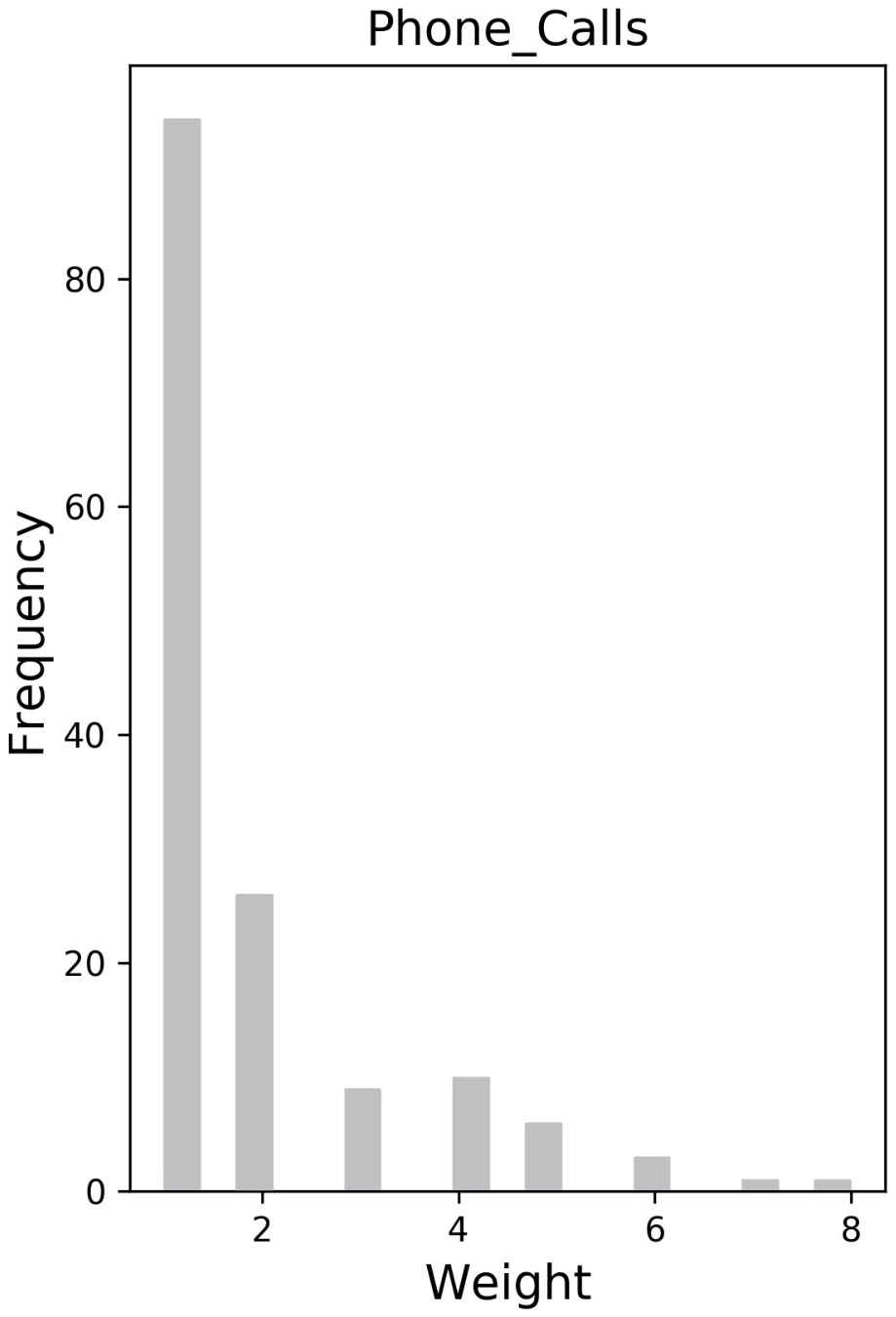}
    \vspace{-2em}
    \caption{\textsc{Phone Calls}}
  \end{subfigure}
    \vspace{-0.8em}
     \caption{Weight Distribution in the \textsc{Meetings} dataset (a), and the \textsc{Phone Calls} dataset (b). (Reproduced from Ficara 
     \emph{et al.} 2020)}
\label{fig:weights}
\end{figure*}

Fig.~\ref{fig:weights} shows the weight distribution of the {\em Meetings} and the {\em Phone Calls} networks. As already mentioned, the weights represent the amount of meetings and phone calls exchanged between pairs of individuals in the networks, respectively. 

It is noteworthy that in both these networks there are just a few high-weight edges {\it i.e.}; nodes incident on those links exhibit an high number of interaction within the network. In~\cite{Ficara2020}, we motivated this behaviour as a necessity from affiliates to focus their efforts in trying to reduce the risk of being intercepted by external people (i.e., LEA, and other people outside the clan). In the {\em Meetings} network, this trend is even more accentuated; moreover, the maximum interactions weight (i.e., $w = 10$) is greater than its counterpart in the {\em Phone Calls} network (i.e., $w = 8$). 

Our explanation is that mobsters prefer to communicate by face-to-face meetings, rather than calling each other, to reduce interception risks. Furthermore, bosses often have to participate to public events to pursue their power inside a clan, including: funerals of other affiliates, and other solemn religious demonstrations (masses, processions, etc.). It is a well known practice that, during those kinds of events, bosses pass messages to their closest subordinate affiliates.

\subsection{Shortest Path Length Analysis}
\label{subsec:path}

\begin{figure*}[htb]
  \centering  
  \begin{subfigure}{.49\textwidth}
    \includegraphics[width=\textwidth]{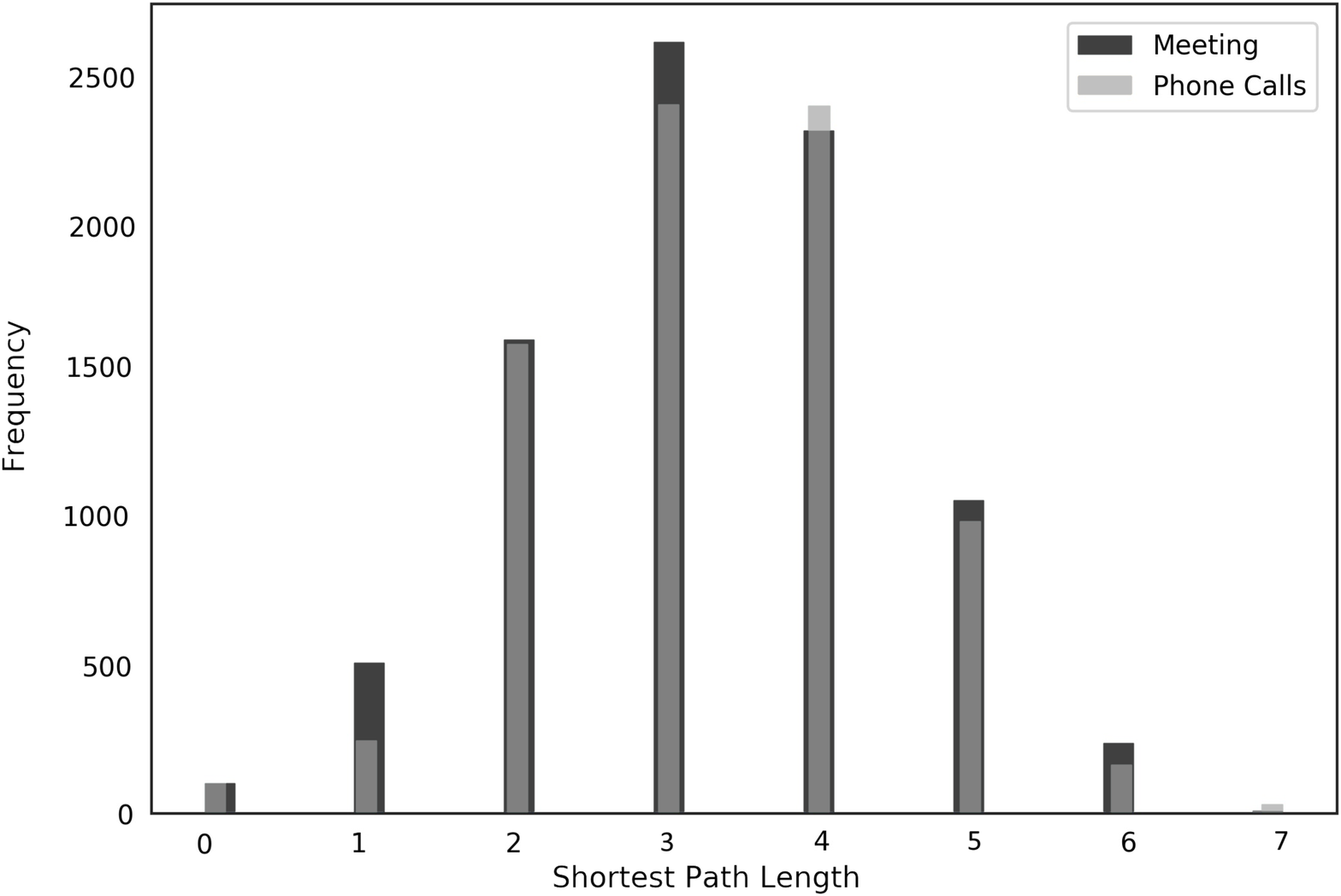}
    \vspace{-2em}
    \caption{\textit{Unweighted} graphs.}
  \end{subfigure}
  \begin{subfigure}{.49\textwidth}
    \includegraphics[width=\textwidth]{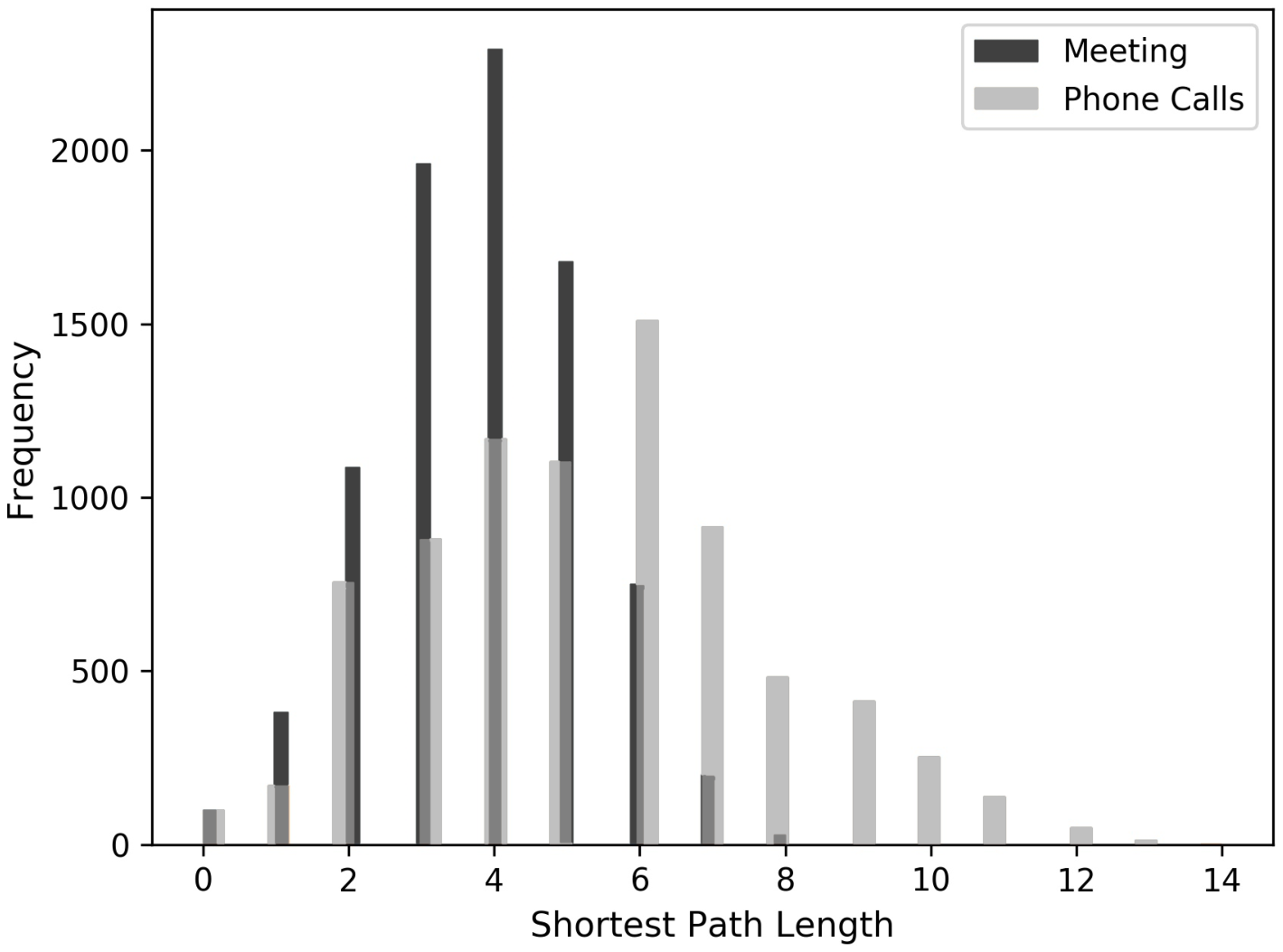}
    \vspace{-2em}
    \caption{\textit{Weighted} graphs.}
  \end{subfigure}
    \vspace{-0.8em}
    \caption{Distribution of shortest path lengths in \textsc{Meetings} and \textsc{Phone Calls} networks in unweighted (a) and weighted (b) graphs. (Reproduced from Ficara \emph{et al.} 2020)}
\label{fig:path}
\end{figure*}

The shortest path length distribution in Fig.~\ref{fig:path} is closely related to dynamic properties such as velocity of messages spreading process within the network. Generally speaking, the criminal organizations structure aims to to optimize the interaction frequency among members, while reducing as much as possible the interception risks. Thus, trusted members may be discovered by following short interactions paths; indeed, those affiliates may also be acting as a \textit{bridge} (See Sect.~\ref{subsec:tools}``Connectedness'') to connect distant groups in the network.

In~\cite{Ficara2020} we noticed that both the weighted and the unweighted shortest path length analyses show a higher interaction frequency among affiliates throughout a ``balanced'' number of intermediates. This means that they do not like to spread their encrypted messages with a too low (resp., high) number of intermediates. This is to avoid, from one side, to overexpose their bosses to police investigations. From the other side, the longer the sequence of intermediates, the higher the chances to be intercepted by people outside the Family.

Even through this analysis, as it was for the weights distribution, it emerged that the clan tries to minimize the risk of interceptions, especially to avoid exposing those mobsters who are hierarchically in a higher rank.

\subsection{Degree Distribution Analysis}
\label{subsec:degdistr}
\begin{figure*}[htb]
  \centering  
  \begin{subfigure}{.49\textwidth}
    \includegraphics [width=\textwidth] {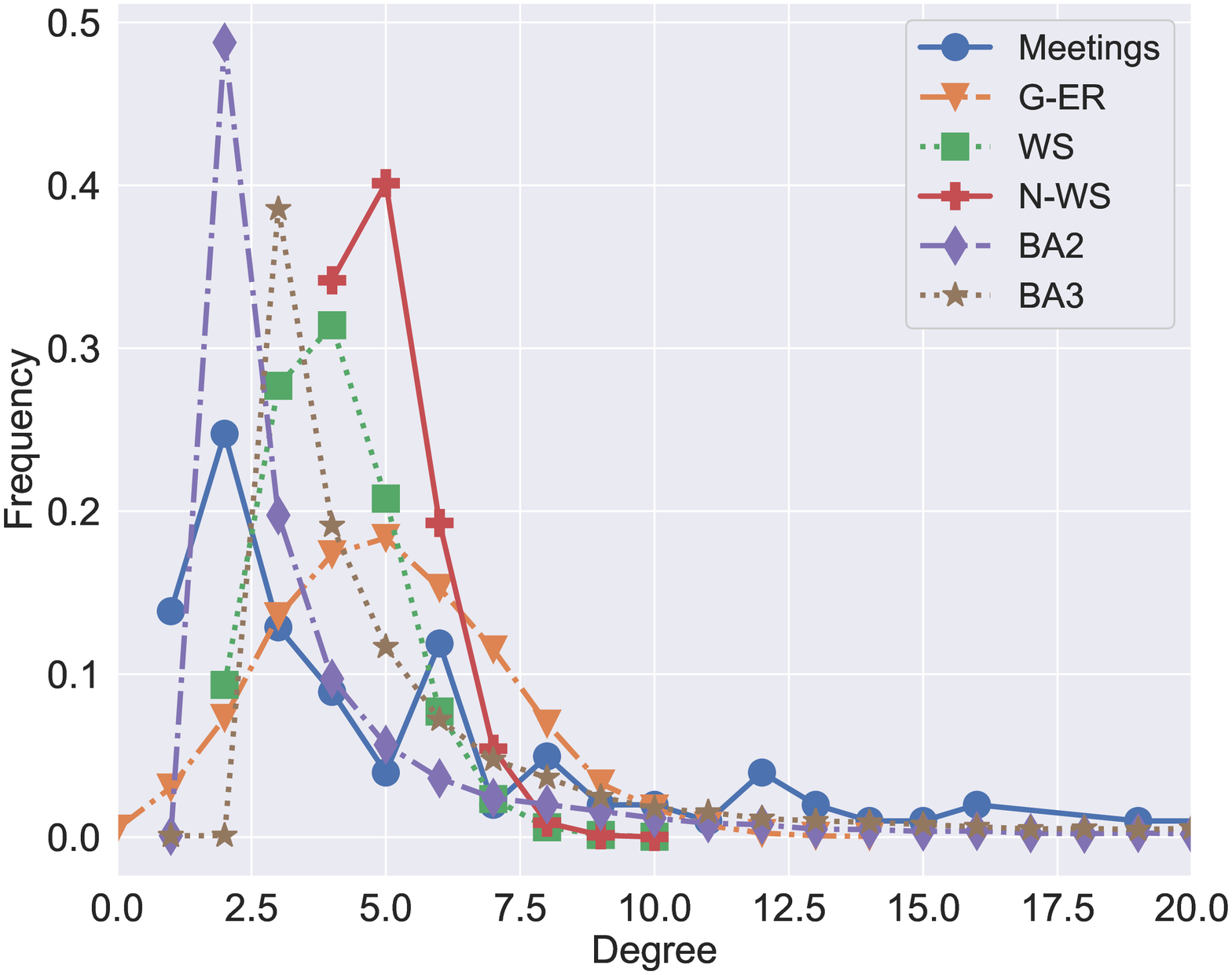}
    \vspace{-2em}
    \caption{\textsc{\textsc{Meetings}}}
  \end{subfigure}
  \begin{subfigure}{.49\textwidth}
    \includegraphics [width=\textwidth] {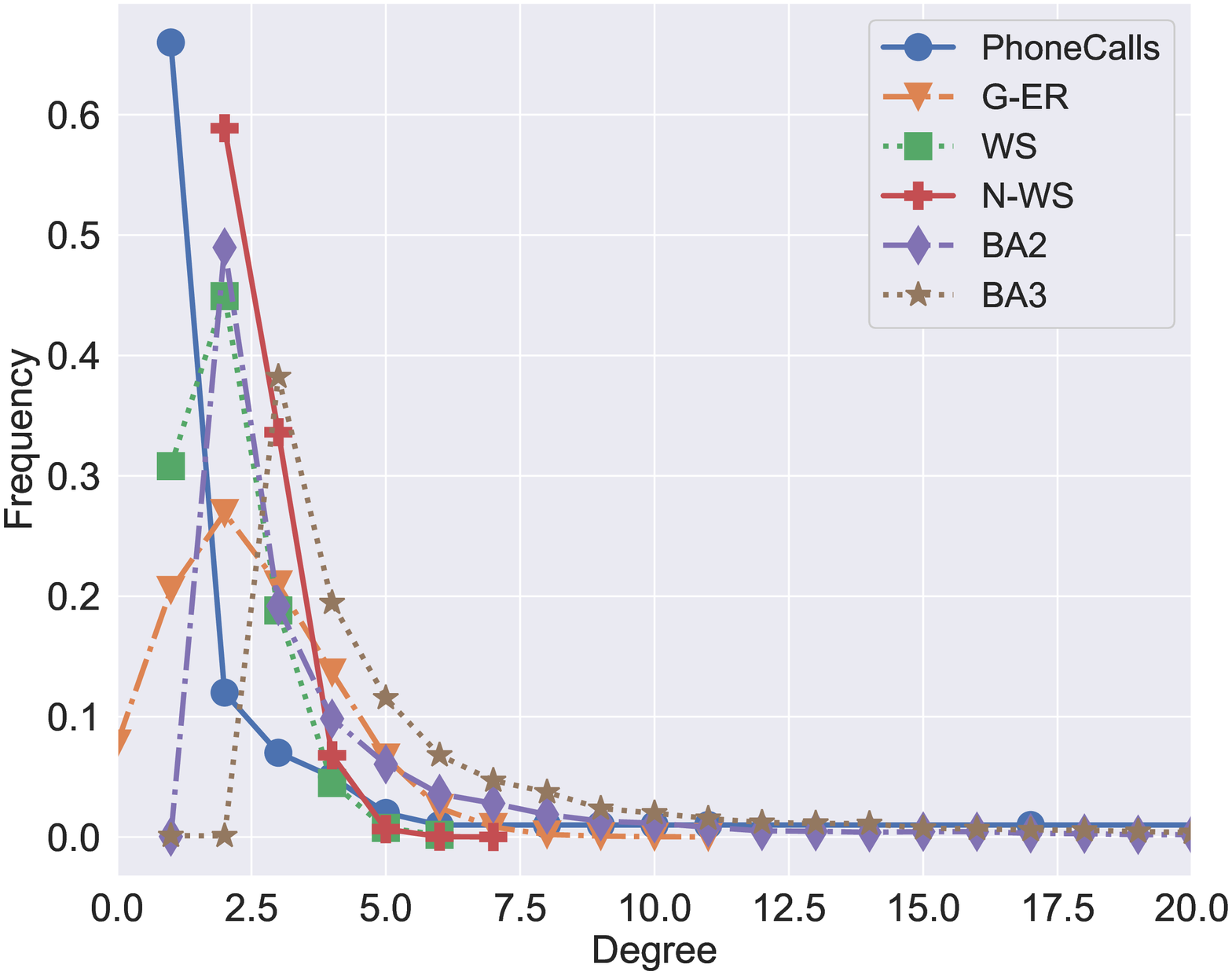}
    \vspace{-2em}
    \caption{\textsc{Phone Calls}}
  \end{subfigure}
    \vspace{-0.8em}
     \caption{Degree Distribution in the \textsc{Meetings} dataset (a), and the \textsc{Phone Calls} datasets (b). Circles give the actual datasets values. G-ER is the Random Network proposed by Gilbert. WS is the Watts-Strogatz network. N-WS is the Newmann variation of WS. BA2 and BA3 are the Barab\'{a}si-Albert models with $m=2$ and $m=3$, respectively.}
\label{fig:degree-distribution}
\end{figure*}

The Degree Distribution Analysis has been conducted in order to discover an appropriate artificial network that would virtually mirror the real-world criminal graphs topology under scrutiny. We previously analysed the weight distribution, but due to lack of libraries available for weighted graphs analysis~\footnote{https://networkx.github.io/documentation/networkx-1.9/reference/generators.html}, we opted for a preliminary analysis on nodes degree distribution. To this end, in Fig.~\ref{fig:degree-distribution} we compared our real criminal networks against five artificial models: \begin{enumerate*}[label=(\roman*)] \item the Random Network by Gilbert (G-ER), \item Watts-Strogatz (WS), \item its variant accordingly with Newmann~\cite{NEWMAN1999341} (N-WS), and \item two different configurations of the Barab\'{a}si-Albert (BA) model in terms of links added at each step; BA2 with $m=2$, and BA3 with $m=3$.
\end{enumerate*} Table~\ref{table:avg-deg-phonecalls} and Table~\ref{table:avg-deg-meetings} summarize the number of edges and average degree obtained with the configurations above described in \textit{Phone Calls} and \textit{Meetings} graphs, respectively.

Note that all the results herein shown represent the average results obtained after 100 runs per each synthetic network.

We initially compared the real networks with the Erd\H{o}s-R\'{e}nyi (ER) topology, but this model did not allow to customize the number of links. Then, we opted for the Gilbert one, whereby both number of nodes $n$ and links $m$ are defined a priori. 

In the WS model, we set $n$, $k=\frac{2m}{n}$ (that represents the number of nearest neighbors links per node), and the rewiring probability $p=0.5$, with $p\in [0,1]$. As previously asserted in Sect.~\ref{subsec:artificial}, if $p=1$, we turn into a Random Network.
The main difference between WS and N-WS models is that in WS, number $p$ is the probability of rewiring each edge; whereas in N-WS, $p=0.25$ is the probability of adding a new edge for each edge. Indeed, if $p=1$, then number of edges is doubled.

\begin{table}[!t]
\caption{Characteristics of Artificial Models in the {\em Phone Calls} network.}
\label{table:avg-deg-phonecalls}
\begin{tabular}{p{3cm}p{2cm}p{2cm}p{2cm}}
\hline\noalign{\smallskip}
\textbf{Model} & No. Edges & Avg. Degree \\
\noalign{\smallskip}\svhline\noalign{\smallskip}
G-ER &  124 & 2.48\\ 
WS &  100 & 2.00\\ 
N-WS &  123 & 2.46\\ 
BA2 &  196 & 3.92\\ 
BA3 &  291 & 5.82\\ 
\noalign{\smallskip}\hline\noalign{\smallskip}
\end{tabular}
\end{table}

\begin{table}[!t]
\caption{Characteristics of Artificial Models in the {\em Meetings} network.}
\label{table:avg-deg-meetings}
\begin{tabular}{p{3cm}p{2cm}p{2cm}p{2cm}}
\hline\noalign{\smallskip}
\textbf{Model} & No. Edges & Avg. Degree \\
\noalign{\smallskip}\svhline\noalign{\smallskip}
G-ER &  256 & 5.07\\ 
WS &  202 & 4.00\\ 
N-WS &  250 & 4.95\\ 
BA2 &  198 & 3.92\\ 
BA3 &  294 & 5.82\\ 
\noalign{\smallskip}\hline\noalign{\smallskip}
\end{tabular}
\end{table}

The actual graphs degree distributions act very differently from one another. In particular, the fluctuations in \textit{Meetings} are justified by the fact that face-to-face encounters have been observed not only between couple of suspects, but also among groups of more than two people at the same time. On the other hand, phone calls have only been considered between individual suspects. 
The analysis suggests that through degree distribution it was not possible to identify an appropriate artificial network that best fits the network characteristics of the two real-world datasets considered. This is mainly due to the size of the networks. In fact, artificial networks seem to work better with lager sizes; thus, are quite unstable in the first step of their creation. 
For example, the emergence of hubs in BA models cannot be highlighted because of the small size of the overall network obtained.

\section*{Conclusions}
\label{sec:conclusions}
This chapter aims to showcase the applicability of Graph Theory in Criminology, during a time when the use of SNA by LEAs is growing substantially. The case study herein reported as an example, explores different approaches on criminal networks analysis by means of network science tools. 

In our study we have first created a graph from data extracted from juridical acts; then, we started a twofold preliminary investigation: a weight distribution analysis, and in parallel, a shortest path length analysis. These have been conduced to identify the extent by which weighted graphs are useful in those small networks. 

Thus, we conducted a comparative degree distribution analysis between our real-world networks and some models generated by popular artificial networks. The aim was to identify the appropriate synthetic network which could simulate criminal networks artificially, but in an effective manner. The strength of this idea is that we may also be able to understand the patterns followed by criminals to create their internal interconnections among affiliates. Our study has found that the network size is a limitation. Indeed, there are significant fluctuations and through degree distribution comparative analysis it was not possible to find an appropriate artificial network that accurately mirrors the two real criminal networks used in our tests. 

To overcome this issue, in future studies we will investigate adjacency matrix structures for both real and synthetic networks to get insights into network topologies.

\end{document}